\documentclass[prd,twocolumn,amsmath,amssymb,floatfix,superscriptaddress,nofootinbib]{revtex4-1}

\usepackage{graphicx}
\usepackage{scrextend}
\usepackage{amssymb}
\usepackage{amsmath}
\usepackage{bm}
\usepackage{color}
\usepackage{multirow}
\usepackage{cprotect}

\DeclareMathAlphabet\mathbfcal{OMS}{cmsy}{b}{n}
\definecolor{darkgreen}{RGB}{50,150,0}
\definecolor{purple}{cmyk}{0.5,1.0,0,0}

\def\edth{\;\raise1.0pt\hbox{$'$}\hskip-6pt\partial}
\def\baredth{\;\overline{\raise1.0pt\hbox{$'$}\hskip-6pt
\partial}}

\def\be{\begin{equation}}
\def\ee{\end{equation}}
\def\ben{\begin{equation} \nonumber}
\def\een{\end{equation}}
\def\ban{\begin{eqnarray*}}
\def\ean{\end{eqnarray*}}
\def\ba{\begin{eqnarray}}
\def\ea{\end{eqnarray}}
\def\({\left(}
\def\){\right)}

\newcommand{\PP}{{\phi\phi}}
\newcommand{\LCDM}{{$\Lambda$CDM }}

\definecolor{ultramarine}{rgb}{0.07, 0.04, 0.56}
\definecolor{cadmiumgreen}{rgb}{0.0, 0.42, 0.24}
\definecolor{indigo(dye)}{rgb}{0.0, 0.25, 0.42}
\usepackage[linktocpage=true]{hyperref}
\hypersetup{
colorlinks=true,
citecolor=ultramarine,
linkcolor=cadmiumgreen,
urlcolor=indigo(dye),
pdfauthor={},
pdftitle={},
pdfsubject={}
}

\begin{document}

\title{Lensing-like tensions in the Planck legacy release}

\author{Pavel Motloch}
\affiliation{Canadian Institute for Theoretical Astrophysics, University of Toronto, M5S 3H8, ON, Canada}

\author{Wayne Hu}
\affiliation{Kavli Institute for Cosmological Physics, Department of Astronomy \& Astrophysics, Enrico Fermi Institute, University of Chicago, Chicago, Illinois 60637, U.S.A}

\begin{abstract}
\noindent
We analyze the final release of the Planck satellite data to constrain the gravitational
lensing potential in a model-independent manner.  The amount of lensing determined
from the smoothing of the acoustic peaks in the temperature and polarization power spectra
is 2$\sigma$ too high when compared with the measurements using the lensing
reconstruction and 2.8$\sigma$ too high when compared with $\Lambda$CDM
expectation based on the ``unlensed'' portion of the temperature and polarization power
spectra. The largest change from  the previous data release is the $\Lambda$CDM expectation, driven by improved
constraints to the optical depth to reionization.  The
anomaly still is inconsistent with actual gravitational lensing, given that the lensing
reconstruction constraints are discrepant independent of the model.
Within the context of $\Lambda$CDM, improvements in its parameter constraints from 
lensing reconstruction bring this tension to  2.1$\sigma$ and from further adding  baryon acoustic oscillation and Pantheon supernova data to a marginally higher 2.2$\sigma$.
Once these other measurements are included, marginalizing
this lensing-like anomaly cannot substantially resolve  tensions with low-redshift
measurements of $H_0$ and $S_8$ in $\Lambda$CDM, $\Lambda$CDM+$N_\mathrm{eff}$ or
$\Lambda$CDM+$\sum m_\nu$; furthermore the  artificial strengthening of
constraints on $\sum m_\nu$ is less than 20\%.
\end{abstract}

\maketitle

\section{Introduction}
\label{sec:intro}

The standard cosmological model, $\Lambda$CDM, has been successful in describing
a large number of diverse cosmological observations. Despite this phenomenological success,
there are several hints \LCDM might not be the end of the story. Most notable is the
difference, statistically significant at over 5$\sigma$, between the values of the Hubble
constant directly measured in the low-redshift Universe \cite{Riess:2019cxk, Wong:2019kwg}
and indirectly extrapolated within the context of $\Lambda$CDM from the observations
sensitive to the early Universe physics \cite{Aghanim:2018eyx, Addison:2017fdm} (see
however \cite{Freedman:2019jwv, Yuan:2019npk}).
The tension between measurements of the amount of clustering in the late
Universe from weak lensing observations \cite{Abbott:2017wau, Chang:2018rxd,
Hikage:2018qbn, Hildebrandt:2018yau, Joudaki:2019pmv} and from the Planck satellite
\cite{Aghanim:2018eyx} is of a more modest statistical significance, but
has persisted over time. Given these differences between experimental constraints, it is
crucial to search for possible systematic effects that could bias our inferences from
individual observations.

Regarding the Planck satellite, there are several ``curiosities'', such as difference
in cosmological parameters when comparing constraints from different ranges of angular
scales \cite{Addison:2015wyg, Aghanim:2016sns} and the large scale anomalies (see
\cite{Akrami:2019bkn} for an overview), that could potentially shed some light on the
detected discrepancies.

This work is motivated by another such ``curiosity'' in the Planck data, the
so-called oscillatory residuals between the Planck temperature power spectra and the
best-fit \LCDM model, visible between angular multipoles $\ell$ of roughly 900 and 1700 \cite{Aghanim:2016sns,
Obied:2017tpd}. In this range of angular scales, these residuals are roughly of opposite
phase to
the CMB peaks and are thus somewhat degenerate with the effects of gravitational
lensing. The amount of gravitational lensing inferred from the
temperature and polarization power spectra is then anomalously high and this
affects constraints on cosmological parameters.  {In this context, the oscillatory
residual curiosity is usually referred to as the $A_L$ or lensing anomaly in the power spectra.}
{Various aspects of this anomaly have been extensively studied \cite{Couchot:2015eea,
Munoz:2015fdv,Valiviita:2017fbx,DiValentino:2019qzk,Raveri:2018wln,DiValentino:2017zyq,
Addison:2015wyg,Heavens:2017hkr,Grandis:2016fwl,Lin:2017bhs,DiValentino:2015ola,Hu:2015rva,
Aghanim:2016sns,Renzi:2017cbg,DiValentino:2015bja}.}

In \cite{Motloch:2018pjy}, we used the 2015 release of the Planck data
\cite{Aghanim:2015xee} to investigate this anomalously high lensing power in depth within
the framework of model-independent lensing constraints. We 
allowed gravitational lensing potentials beyond the $\Lambda$CDM model and studied
implications of such deviations. Recently, the legacy version of the Planck
collaboration products was released, with large and small scale
polarization data finally deemed to be sufficiently systematics free to
be useable for a cosmological analysis. It is thus timely to reevaluate our previous
analysis with the new data and make definitive model-independent statements about the gravitational lensing
constraints from Planck, including a final assessment of the significance of various
tensions found in the data. We also show how fully marginalizing parameters
associated with the lensing anomaly
in the power spectra affects selected cosmological parameters of
interest in \LCDM and two of its extensions, \LCDM+$N_\mathrm{eff}$ and \LCDM+$\sum
m_\nu$.

This paper is organized as follows: We start by summarizing the data and likelihoods used in this
work in \S~\ref{sec:data}. In \S~\ref{sec:lensPCs} we briefly review the technique for probing
the gravitational lensing potential in a model-independent fashion and in
\S~\ref{sec:mcmc} provide details of our Markov Chain Monte Carlo (MCMC) analysis.
\S~\ref{sec:lensing} presents lensing constraints derived from the Planck likelihoods and
\S~\ref{sec:parameters} assesses the effects the Planck lensing anomaly has
on constraints of cosmological parameters. We conclude with discussion in
\S~\ref{sec:discuss}.  In the Appendix, we justify our choice of tension statistic
with a multidimensional analysis.

\section{Data and likelihoods}
\label{sec:data}

To constrain values of cosmological parameters, we mainly use likelihoods derived from 
data collected by the Planck satellite. All Planck
likelihoods used in this work are listed in Table~\ref{tab:planck_likes}.
As usual, $T$ stands for temperature, $E$ and $B$ for the polarization modes and $\phi$
for the gravitational lensing potential.

We primarily use the legacy versions of the temperature, polarization and lensing likelihoods
\cite{Aghanim:2019ame, Aghanim:2018oex}. To better constrain cosmological parameters,
especially $\tau$, small scale temperature and polarization likelihoods are always
accompanied by the large scale TT and EE likelihoods. We also investigate how some of our results
change when we replace the official Planck large scale EE likelihood by its post-legacy
version based on a reanalysis of the Planck High Frequency Instrument (HFI) data 
\cite{Pagano:2019tci}.

We compare the new constraints with their counterparts based on the likelihoods included in
the second Planck data release \cite{Aghanim:2015xee,Ade:2015zua}. This older release 
included information from the large scale TE and BB power spectra which we do not use in
the legacy analysis, but this does not significantly affect constraints on the
gravitational lensing potential.

\begin{table*}
\caption{Planck likelihoods used in this work} 
\label{tab:planck_likes}
\begin{tabular}{ccccc}
\hline\hline
Label & Release & Power spectra & $\ell$-range & Name\\
\hline
\verb|TT| & legacy & TT & $\ell \ge 30$ & \verb|plik_rd12_HM_v22_TT|\\
	& 2015 & TT & $\ell \ge 30$ &\verb|plik_dx11dr2_HM_v18_TT|\\
\verb|TTTEEE| & legacy & TT,TE,EE & $\ell \ge 30$ & \verb|plik_rd12_HM_v22b_TTTEEE|\\
	& 2015 & TT,TE,EE & $\ell \ge 30$ & \verb|plik_dx11dr2_HM_v18_TTTEEE|\\
\verb|lowl| & legacy & TT,EE & $\ell < 30$ & \verb|commander_dx12_v3_2_29|,\\
	&&&& \verb|simall_100x143_offlike5_EE_Aplanck_B|\\
	& 2015 & TT,TE,EE,BB & $\ell < 30$ & \verb|lowl_SMW_70_dx11d_2014_10_03_v5c_Ap|\\
\verb|sroll2| & post-legacy & TT,EE & $\ell < 30$ & \verb|commander_dx12_v3_2_29|,\\
	&&&& \verb|simall_100x143_sroll2_v3_EE_Aplanck|\\
\verb|PP| & legacy & $\PP$ & $8 \le \ell \le 400$ & \verb|smicadx12_Dec5_ftl_mv2_ndclpp_p_teb_consext8_CMBmarged|\\
	& 2015 & $\PP$ & $40 \le \ell \le 400$& \verb|smica_g30_ftl_full_pp|\\
\hline\hline
\end{tabular}
\end{table*}

For certain analyses we additionally use baryon acoustic oscillation likelihoods
(\verb|BAO|) from \cite{Alam:2016hwk,Beutler:2011hx,Ross:2014qpa}, Pantheon supernova
likelihood (\verb|SN|) \cite{Scolnic:2017caz} and the South Pole Telescope (SPT) TE and EE
power spectra likelihood \cite{Henning:2017nuy}.

\section{Lensing parameterization}
\label{sec:lensPCs}

In this work we apply a technique to directly constrain the CMB gravitational lensing
potential from the CMB data in a model-independent
way. In this section we provide a brief overview of this technique, while more details can be
found in \cite{Motloch:2016zsl,Motloch:2017rlk}.

\begin{figure}[b]
\center
\includegraphics[width = 0.49 \textwidth]{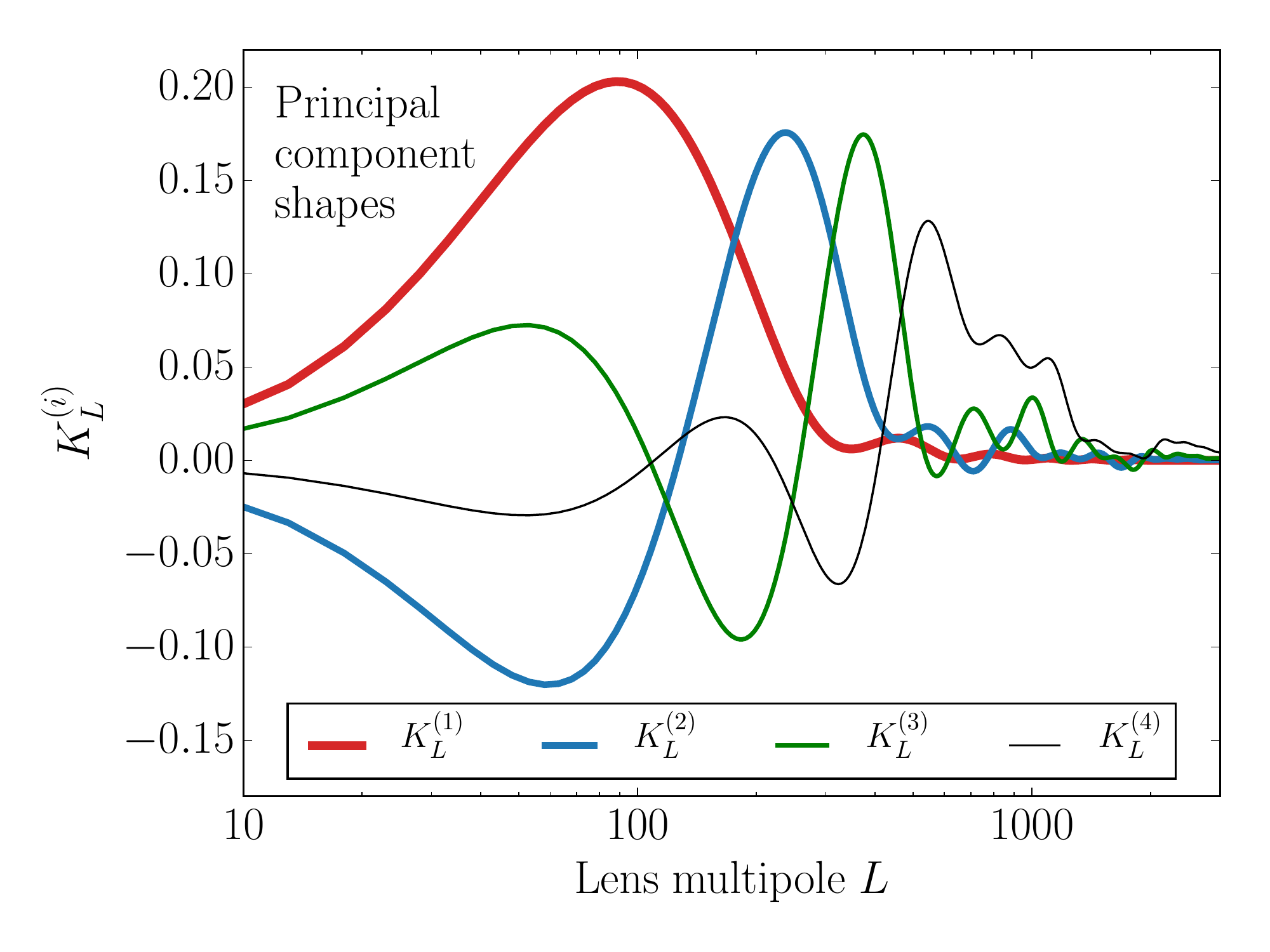}
\caption{Lensing principal components $K_L^{(i)}$ used in this work.
}
\label{fig:pcs} 
\end{figure}

We parameterize the gravitational lensing potential power spectrum in terms of 
$N_\mathrm{pc}$ effective parameters $\Theta^{(i)}$, which determine arbitrary
variations around a fixed fiducial power spectrum $C_{L, \mathrm{fid}}^\PP$ as 
\be
\label{definition}
	C_L^\PP = C_{L, \mathrm{fid}}^\PP \exp\(\sum_{i = 1}^{N_\mathrm{pc}} K^{(i)}_L\, \Theta^{(i)} \) .
\ee
In this setup, constraining $\Theta^{(i)}$ from the data corresponds directly to
constraining the gravitational lensing potential. This should be contrasted with the
common approach of introducing a phenomenological parameter $A_L$ which multiplies
$C_L^{\phi\phi}$ at each point in the model space and cannot be  interpreted {in terms of the
lensing potential} once model
parameters are marginalized over.

To allow easier comparison with our previous results, we use the same $C_{L,
\mathrm{fid}}^\PP$ and $K_L^{(i)}$ as in \cite{Motloch:2018pjy}.
The fiducial cosmological model used to calculate $C_{L, \mathrm{fid}}^\PP$ in
Eq.~\eqref{definition} is taken from the best fit flat $\Lambda$CDM cosmological model,
determined from the Planck 2015 \verb|TT+lowl| likelihoods, with no primordial tensor
modes and minimal mass neutrinos ($\sum m_\nu = 60\, \mathrm{meV}$).

Table~\ref{tab:fiducial} lists values of the corresponding cosmological parameters;
the symbols have their standard meaning: $\Omega_b h^2$ is the physical baryon density; $\Omega_c
h^2$ the physical cold dark matter density; $n_s$ the tilt of the scalar power spectrum;
$\ln A_s$ its log amplitude at $k=0.05$ Mpc$^{-1}$; $\tau$ the optical depth through
reionization, and $\theta_*$ the angular scale of the sound horizon at recombination. 
To reflect the best constraints on $\tau$ at the time,
its value was taken from \cite{Adam:2016hgk} and $A_s$ decreased to keep $A_s
e^{-2\tau}$ constant.

While a lower $A_s$ tends to exacerbate the preference for anomalously high lensing in the
temperature power spectrum within the $\Lambda$CDM context, here it serves only to define
the baseline fiducial model against which $\Theta^{(i)}$ are measured and does not affect 
the significance of tensions discussed below.
{If we used a different fiducial model to define $C_{L, \mathrm{fid}}^\PP$, for
example one based on the Planck 2018 data, each $\Theta^{(i)}$ parameter would be
shifted by a constant, the value of which would depend on the ratio of the two fiducial
$C_L^\PP$. For example, in Fig.~\ref{fig:2d_contours_ttteee} this would lead to a coherent shift of the
contours relative to the origin, but the tensions between datasets would remain unchanged
due to the coherent nature of this shift.
}

\begin{table}
\caption{$\Lambda$CDM parameters and their fiducial values 
for the lens PC construction\footnote{In $\Lambda$CDM, these parameters also imply a Hubble constant
of $h=0.6733$.}.
}
\label{tab:fiducial}
\begin{tabular}{c|c}
\hline\hline
Parameter & Fiducial value\\
\hline
100 $\theta_*$ & 1.041 \\
$\Omega_c h^2$ & $0.1197$ \\
$\Omega_b h^2$ & $0.02223$ \\
$n_s$ & $0.9658$\\
$\ln(10^{10} A_s)$ & $3.049$ \\
$\tau$ & $0.058$\\
\hline\hline
\end{tabular}
\end{table}

The $K_L^{(i)}$ in \eqref{definition} are chosen such that $\Theta^{(i)}$ correspond to
the $N_\mathrm{pc}$ principal components (PCs) of the gravitational lensing potential best
measured by Planck 2015 temperature data, as determined using a Fisher matrix construction
\cite{Motloch:2018pjy}. The resultant eigenmodes $K_L^{(i)}$ are shown in
Fig.~\ref{fig:pcs}. We retain $N_\mathrm{pc}=4$ PCs in order to fully characterize all
sources of lensing information \cite{Motloch:2018pjy}.

\section{MCMC analysis}
\label{sec:mcmc}

We use CosmoMC\footnote{https://github.com/cmbant/CosmoMC} \cite{Lewis:2002ah}
to sample the posterior probabilities in the various parameter spaces. Each of our chains
has a sufficient number of samples such that the Gelman-Rubin statistic $R-1$
\cite{Gelman:1992zz} falls below at least 0.01.

We assume flat tophat priors on $\Theta^{(i)}$ and choose an uninformative prior
on $\Theta^{(1)}$, as it is the variable in which we will evaluate the tensions between
data sets. As the lensing reconstruction constraint on $\Theta^{(2)}$ is still somewhat
informative, we restrict $\Theta^{(2)}$ to lie within six standard deviations from its
mean value, both obtained from the Planck lensing reconstruction likelihood. For
consistency with our previous work, this prior is based on the 2015 release of the
likelihood. This prior does not significantly change any of the tensions we find below.
For the remaining two $\Theta^{(i)}$, that allow further freedom in the shape of the
gravitational lensing potential, we limit their variation such that all $C_L^\PP$ are
within a factor of 1.5 of $C_{L,\rm fid}^\PP$. These weak priors are meant to eliminate
cases that would be in conflict with other measurements of large scale structure or imply
unphysically large amplitude high frequency features in $C_L^\PP$. 
Table~\ref{tab:priors} summarizes the ranges over which we allow $\Theta^{(i)}$ to vary.

\begin{table}
\caption{Flat $\Theta^{(i)}$ priors used in this work.}
\label{tab:priors}
\begin{tabular}{c|c}
\hline\hline
Parameter & \phantom{aaa}Prior range\phantom{aaa}\\
\hline
$\Theta^{(1)}$ & [-2.00, 2.00] \\
$\Theta^{(2)}$ & [-2.10, 1.67] \\
$\Theta^{(3)}$ & [-2.32, 2.32] \\
$\Theta^{(4)}$ & [-3.16, 3.16] \\
\hline\hline
\end{tabular}
\end{table}

In analyses that include CMB temperature and polarization power spectra, in addition to
the four lensing parameters $\Theta^{(i)}$ we also vary the six $\Lambda$CDM parameters
(or more in the extensions), with flat uninformative priors. Unlike the standard analysis,
these parameters do not affect the lensing potential $C_L^\PP$ used to calculate the lensed CMB
temperature and polarization power spectra and only affect the unlensed power spectra at
recombination.  
Sometimes, we will refer to these parameters collectively as $\tilde
\theta_A$, where the tilde is to remind the reader that the gravitational lens potential
entering the calculation of the lensed CMB power spectra is not changed by these
parameters but is fully determined by $\Theta^{(i)}$ through \eqref{definition}. 

Given constraints on the parameters $\tilde \theta_A$ and assuming a particular
cosmological model, for instance $\Lambda$CDM, it is possible to make a prediction of the values of $\Theta^{(i)}$,
which then can be compared with the direct constraints. 

In our analysis, the lensing parameters $\Theta^{(i)}$ serve two distinct purposes. In \S~\ref{sec:lensing} we use them 
to {constrain $C_L^\PP$ and} study tensions between various sources of lensing information in the
data, while in \S~\ref{sec:parameters} we use $\Theta^{(i)}$ as nuisance parameters to
marginalize over  and remove the lensing-like anomaly in the temperature and polarization power spectra to uncover its effects on cosmological parameters.
In particular, in the former case \verb|PP|  is used to directly constrain
$\Theta^{(i)}$ through \eqref{definition}. In the latter 
{\verb|PP| depends on the parameters $\tilde\theta_A$ of the
investigated model in the usual manner, which allows us to constrain these parameters better.}
This then indirectly improves lensing parameter constraints from the temperature
and polarization power spectra as we shall see.

We use default foreground and nuisance parameters and their priors in all the
likelihoods.

\section{Lensing constraints from the final Planck release}

\label{sec:lensing}

\begin{figure}
\center
\includegraphics[width = 0.49 \textwidth]{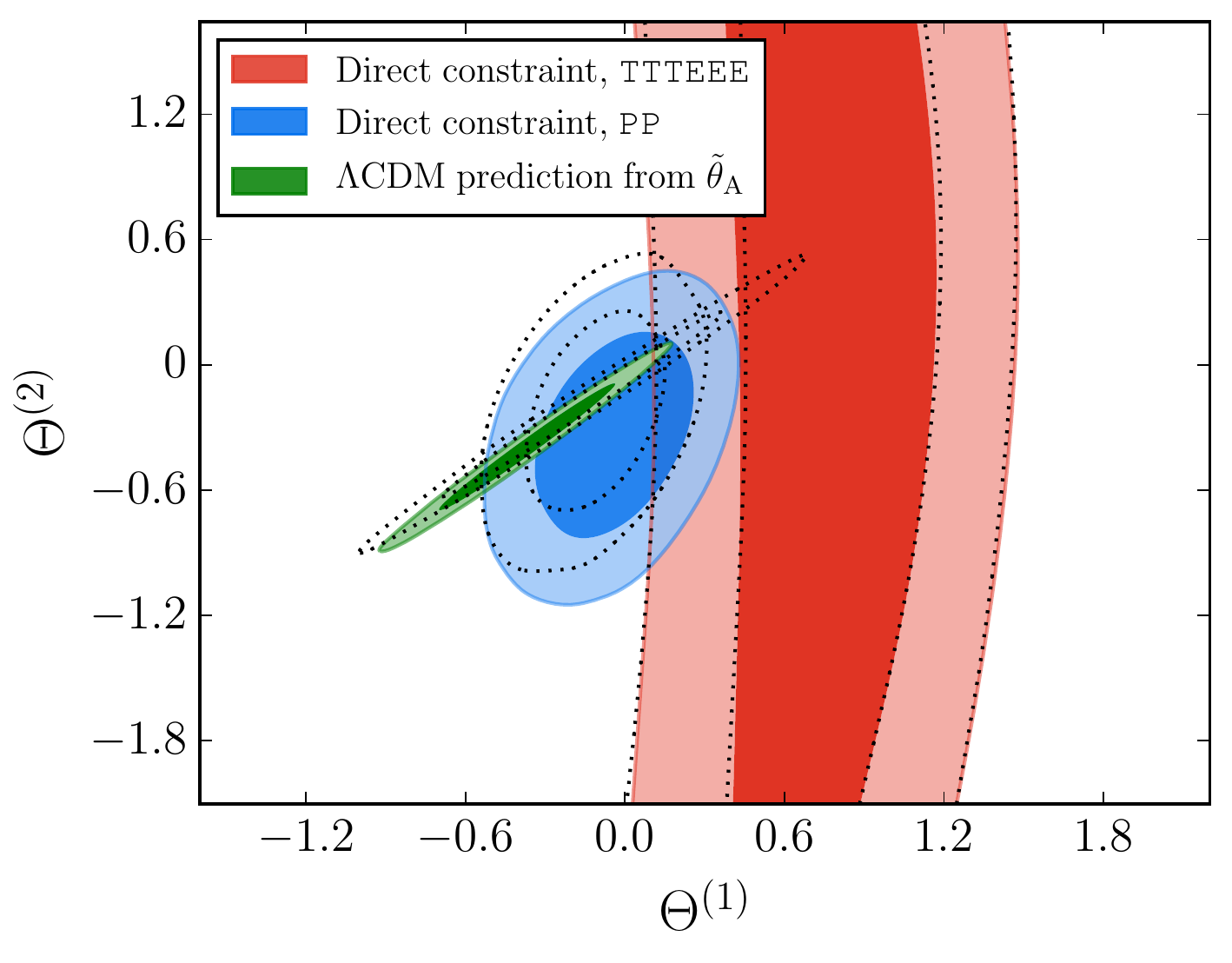}
\cprotect
\caption{CMB power spectrum constraints on lens PCs $\Theta^{(1)}$
and $\Theta^{(2)}$ from \verb|TTTEEE+lowl| (red, 68\% and 95\% CL) compared
with lens reconstruction \verb|PP| (blue) and $\Lambda$CDM predictions based on unlensed
parameters $\tilde \theta_A$ from \verb|TTTEEE+lowl| (green). 
Dashed contours show the constraints when using the 2015 likelihoods. 
}
\label{fig:2d_contours_ttteee}
\end{figure}

\begin{figure}
\center
\includegraphics[width = 0.49 \textwidth]{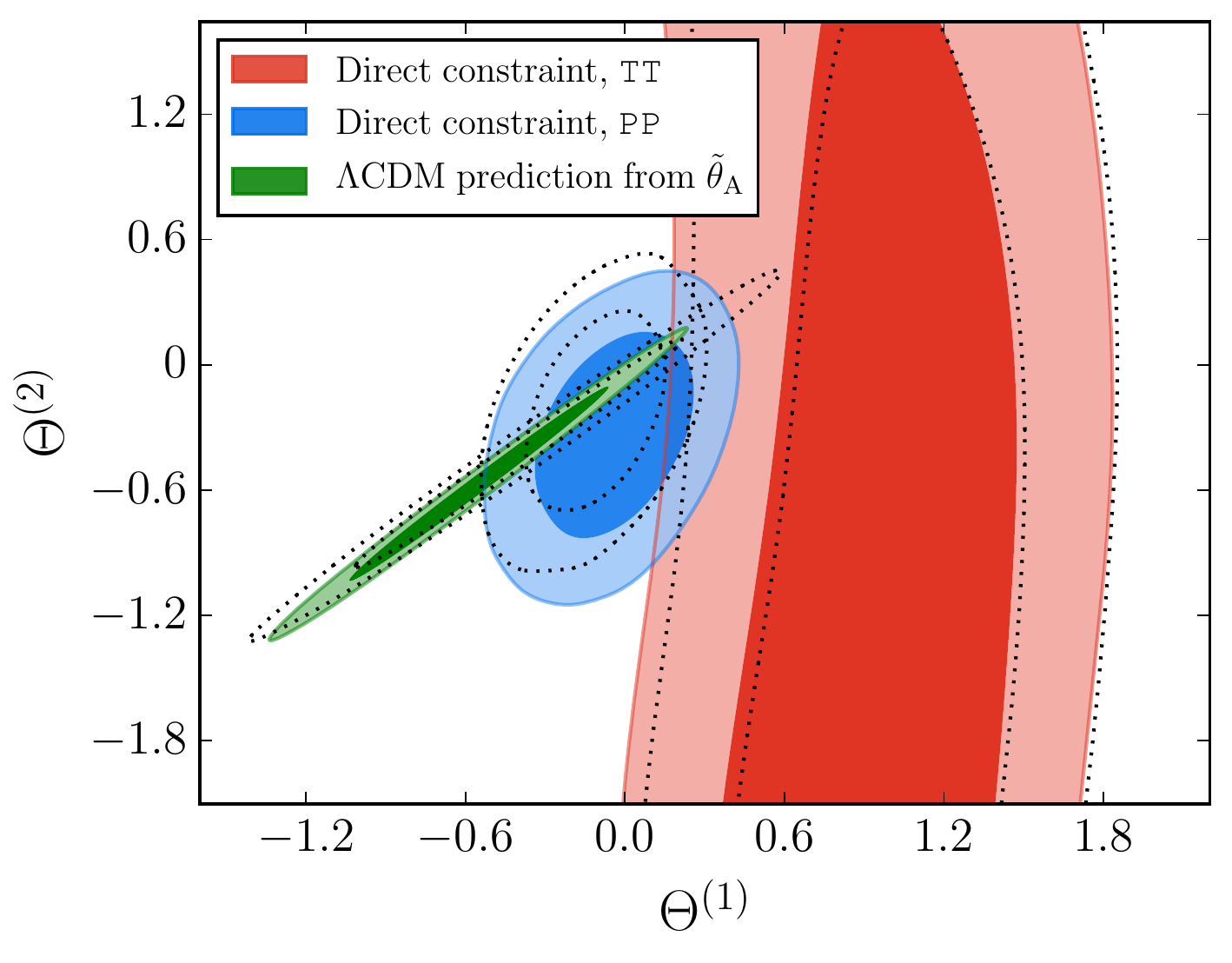}
\cprotect\caption{Same as Fig.~\ref{fig:2d_contours_ttteee}, but with high-$\ell$ polarization data
neglected, i.e. \verb|TTTEEE+lowl| is replaced by \verb|TT+lowl|. 
   }
\label{fig:2d_contours_tt}
\end{figure}

In Fig.~\ref{fig:2d_contours_ttteee} we show constraints on $\Theta^{(1)}, \Theta^{(2)}$
that parameterize the CMB gravitational lensing potential in a model independent manner. We compare here the direct constraints on $\Theta^{(1,2)}$ from the
\verb|TTTEEE+lowl| (red) and
\verb|PP| (blue) likelihoods with the \LCDM predictions (green) based on the
\verb|TTTEEE+lowl| constraints on $\tilde \theta_A$, the standard cosmological parameters
constrained only through their effect on the unlensed power spectra. In
Fig.~\ref{fig:2d_contours_tt} we show the same, with the information from the small scale
polarization dropped, i.e.\ \verb|TTTEEE| replaced by \verb|TT|.

The amount of lensing manifested in the high-$\ell$ temperature and
polarization likelihoods is notably higher than the amount of lensing inferred from the
lensing reconstruction or the one predicted within $\Lambda$CDM as evidenced from the barely
overlapping 95\% confidence level (CL) regions.
As expected, the \verb|TT+lowl| / \verb|TTTEEE+lowl| data can directly
constrain only the leading lensing principal component $\Theta^{(1)}$ and do not provide
competitive constraints on $\Theta^{(2)}$ (or the higher lensing PCs). By construction, the two lensing PCs are also  mostly uncorrelated 
for these two cases (red contours).   As discussed in
\cite{Motloch:2018pyt}, the degeneracy direction apparent in the $\Lambda$CDM predictions 
(green contours) is closely
related to changes in the lensing potential produced by shifts in $A_s$ and $\Omega_c
h^2$, the two parameters on which $C^\PP_L$ mainly depends in $\Lambda$CDM.

In Figs.~\ref{fig:2d_contours_ttteee} and \ref{fig:2d_contours_tt} we use dashed contours
to denote the same probability contours derived using the 2015 Planck data. The direct
constraints from the smoothing of the acoustic peaks (red) do not significantly change, which is
not surprising given only minor changes in the high-$\ell$ Planck likelihoods and insensitivity
of $\Theta^{(1,2)}$ constraints to the details of the low-$\ell$ likelihoods \cite{Motloch:2018pjy}.
The lensing reconstruction constraints (blue) are weakened, because
in the legacy release the Planck ``lensing only'' likelihood adds marginalization over the
uncertainties in the cosmological parameters, that
affect lensing bias subtractions. The largest changes are in the predictions within
the context of the \LCDM model (green); these mostly reflect changes in $\tau$ constraints and
the related breaking of the $A_s e^{-2\tau}$ degeneracy of the high-$\ell$ data.

\begin{figure}
\center
\includegraphics[width = 0.49 \textwidth]{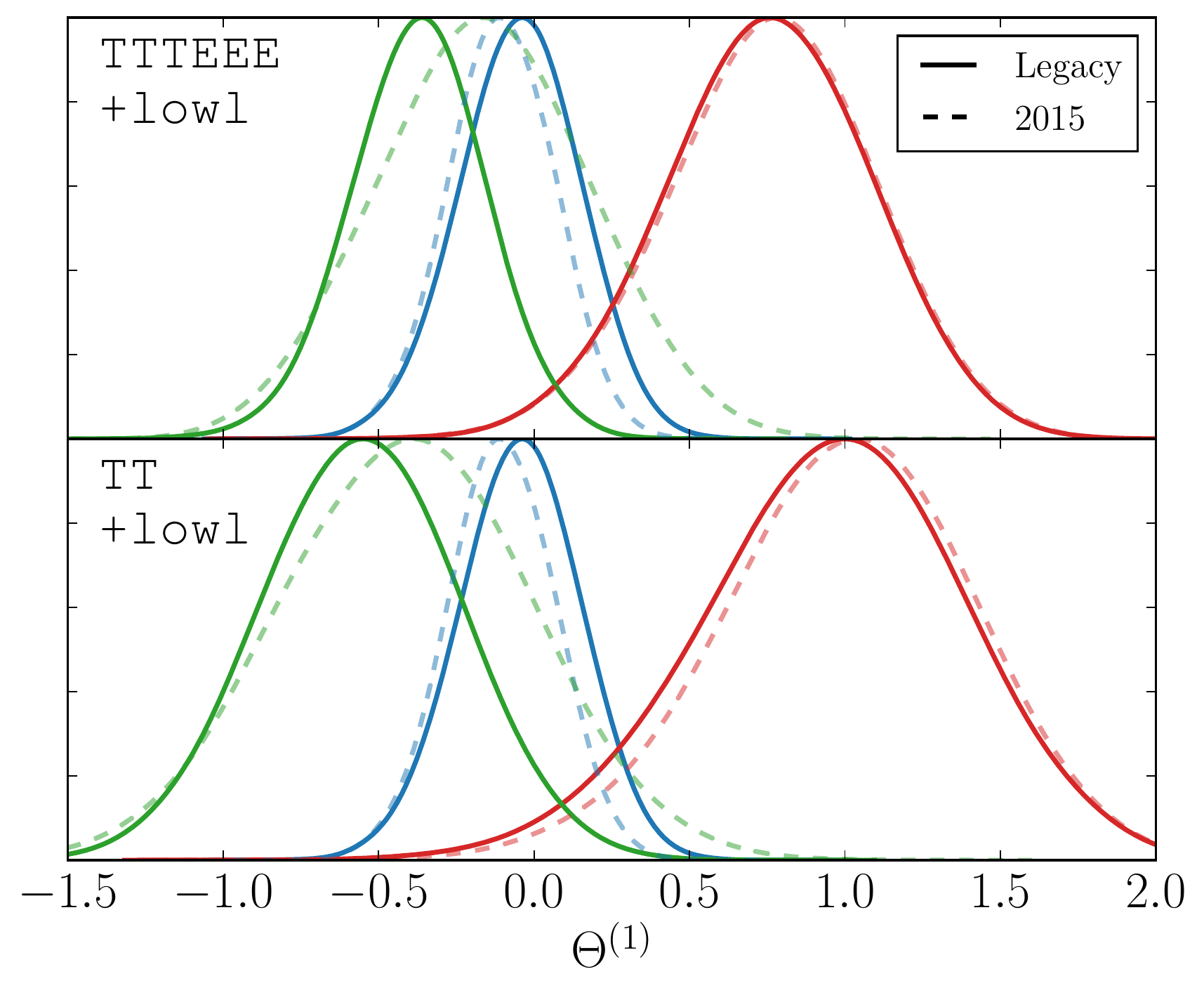}
\cprotect\caption{Posterior probability distributions for $\Theta^{(1)}$.
Direct constraints from \verb|PP| (blue), direct constraints from
\verb|TTTEEE+lowl| (red, top) or \verb|TT+lowl| (red, bottom) and \LCDM
predictions based on constraints on the ``unlensed'' cosmological parameters $\tilde \theta_A$
from \verb|TTTEEE+lowl| (green, top) or \verb|TT+lowl| (green, bottom). Solid curves
correspond to constraints obtained with the Planck legacy likelihoods, dashed
with the 2015 version of the likelihoods.
}
\label{fig:1d}
\end{figure}

In Fig.~\ref{fig:1d} we show the constraints on $\Theta^{(1)}$, marginalized over all the
other parameters. Using these posterior probability distributions, we calculate the
significance of  tensions between the various constraints. Because the posteriors are
non-Gaussian, we use the generalization of the ``shift in the means'' statistic to
non-Gaussian posteriors, introduced in Appendix~C of \cite{Motloch:2018pyt}.
Note that the choice to investigate tensions in $\Theta^{(1)}$, the only such
parameter to be significantly constrained by the temperature and polarization power spectra -- as
opposed to
some function of $\Theta^{(i)}$ with free parameters -- was made a priori and so we do not
have to consider the look-elsewhere effect, despite adding four new parameters.

Comparing the direct constraints from \verb|TTTEEE+lowl| {(red in
Fig.~\ref{fig:1d})} with  {either} the lensing reconstruction
constraints from \verb|PP|  {(blue)} or the \LCDM prediction based on the constraints
on the ``unlensed'' cosmological parameters $\tilde \theta_A$  {(green)}, we find
tensions of 2.0$\sigma$ and 2.8$\sigma$  {respectively. While the latter is closely
related to the usually discussed $A_L$ tension, the former shows that the amount of peak
smoothing in Planck power spectra is somewhat unusual also when compared with the lensing
reconstruction.}  Dropping the high-$\ell$ polarization data changes the respective
tensions to 2.1$\sigma$ and 2.7$\sigma$.

{
In Appendix~\ref{sec:appendix}, we confirm the $\sim 2 \sigma$ tension between the
direct constraints from \verb|TTTTEEE+lowl| and \verb|PP| using update difference in mean statistic developed in
\cite{Raveri:2018wln}. This statistic was developed to investigate tensions that involve a
data set that is constraining only a subset of the parameters (\verb|TTTEEE+lowl|
strongly constraining only $\Theta^{(1)}$ in our case). This result shows that
marginalizing over other $\Theta^{(i)}$ does not artificially degrade significance of the
discovered tension.
}

Given {that} the \verb|TTTEEE+lowl|  and  \verb|PP| constraints are in tension even when allowing arbitrary $C_L^\PP$, while the $\Lambda$CDM prediction is in good agreement 
with the latter, it is not likely that
gravitational lensing is responsible for the lensing-like anomaly.

\begin{figure}
\center
\includegraphics[width = 0.49 \textwidth]{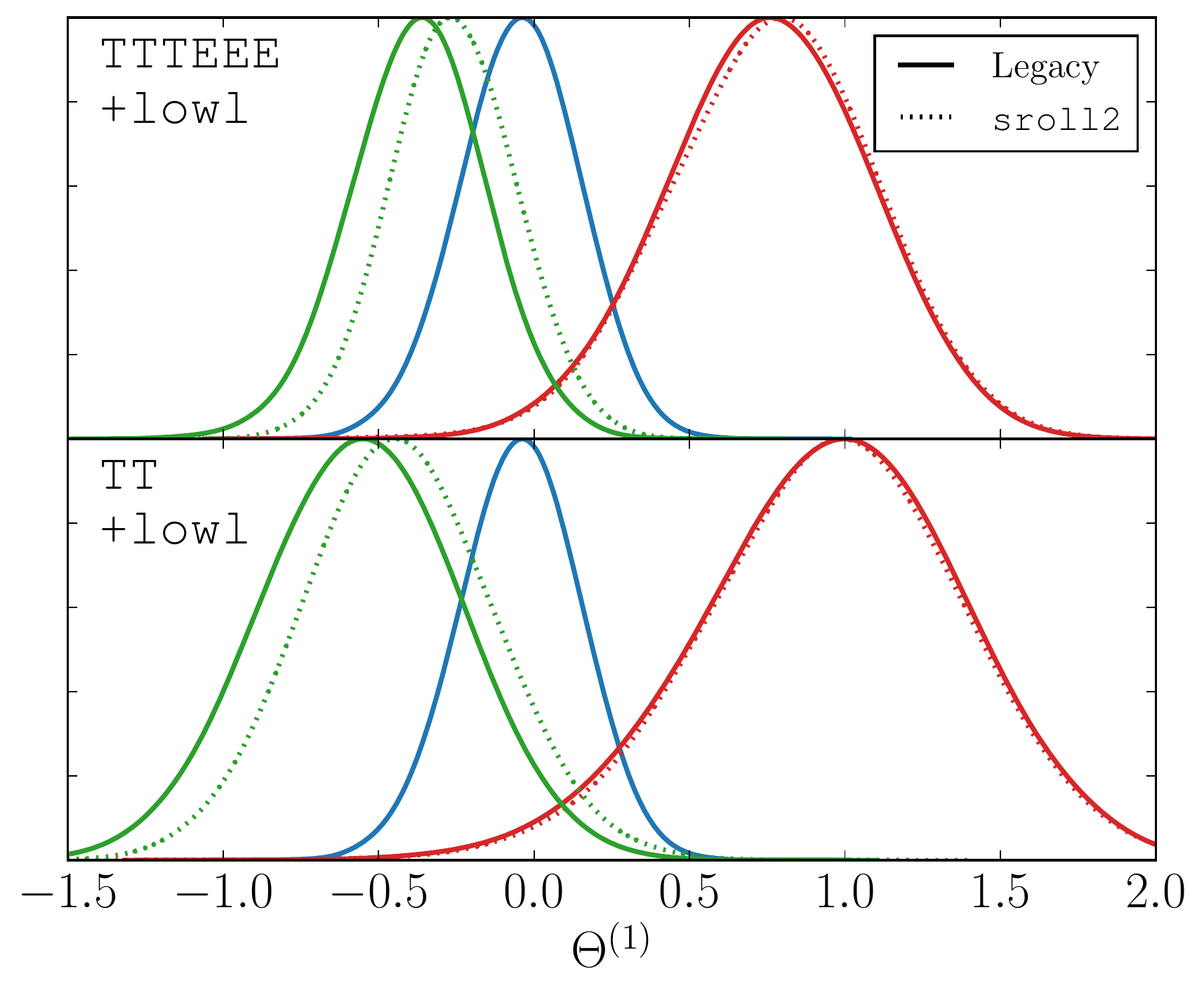}
\cprotect\caption{Effect of the reanalysis of the large scale Planck HFI data. The solid curves as
in Fig.~\ref{fig:1d}, the dotted curves show effects of replacing the official Planck
large scale EE likelihood with the \verb|sroll2| likelihood from \cite{Pagano:2019tci}.
} 
\label{fig:1d_sroll2}
\end{figure}

In Fig.~\ref{fig:1d_sroll2}, we show the effect of replacing the official Planck
low-$\ell$ EE likelihood with \verb|sroll2|,  an improved analysis of the Planck HFI 
polarization
data
 \cite{Pagano:2019tci}. The tension between the direct constraints
on $\Theta^{(1)}$ from the smoothing of the peaks and the \LCDM expectation in this case
decreases to 2.5$\sigma$, both when high-$\ell$ polarization is present or absent. This is
due to an increased best fit value of $\tau$ (see \cite{Pagano:2019tci}
for a similar discussion within the context of the $A_L$ parameter).

Because the direct constraints on the lensing principal components $\Theta^{(i)}$ from the
Planck temperature and polarization power spectra do not notably change, the significances
of the tensions with the SPT data presented in \cite{Motloch:2018pyt}
change by less than 0.1$\sigma$. The tension between lensing constraints from
Planck high-$\ell$ temperature power spectra and lensing constraints from SPT TE and EE
power spectra remains at 3.0$\sigma$, adding Planck polarization decreases the
significance to 2.9$\sigma$.

Related to this, let us comment on a possibility of the Planck lensing-like anomaly being driven by 
new physics at recombination that changes the CMB temperature and
polarization power spectra in a way similar to lensing. If this was the case, we would
expect SPT temperature and polarization data to also show preference for high amount of
lensing. However, these data actually prefers somewhat lower amplitude than the Planck
lensing reconstruction, although with large error bars \cite{Motloch:2018pyt}.

\section{Cosmological parameter  impact}
\label{sec:parameters}

If, despite significant care and effort of the Planck Collaboration, the observed lensing-like
anomaly fully or partially originates in an unknown systematic effect, constraints on
cosmological parameters would be directly affected.
In this section, we comment on how the preference for anomalously high lensing amplitude in
the Planck temperature and polarization power spectra affects constraints on selected
cosmological parameters, especially in the context of parameter tensions with other
experiments.  

In this section, $\Theta^{(i)}$ only enter the $C_L^\PP$ used to calculate the
temperature and polarization power spectra; the \verb|PP| likelihood uses the standard
$C_L^\PP$ of the given cosmological model, evaluated using $\tilde \theta_A$. This way,
marginalizing over $\Theta^{(i)}$ removes any effect of the extra peak smoothing and other
lensing effects in the temperature and polarization power spectra on constraints of the
cosmological parameters.  This generalizes marginalization over $A_L$ that is usually
employed for this purpose.

We study $\Lambda$CDM and two of its extensions, where either
$N_\mathrm{eff}$, the effective number of light relativistic degrees of freedom, or $\sum
m_\nu$, the sum of the neutrino masses, are allowed to vary.

\subsection{\LCDM}

Within the context of $\Lambda$CDM, the anomalously high lensing power in temperature and
polarization power spectra acts toward
increasing the best fit values of $\Omega_c h^2$ and $A_s$. Due to parameter degeneracies,
this leads to related shifts in $\Omega_b h^2, H_0$ and $n_s$. Once
the lensing PCs are allowed to absorb the extra lensing, all the cosmological
parameter readjust accordingly. Notably, $\Omega_c h^2$ shifts from the
$\Lambda$CDM  value $0.1202 \pm 0.0014$  to $\widetilde{\Omega_c h^2} = 0.1179
\pm 0.016$.
This then affects the interpretation of tensions with local measurements.

\begin{figure*}
\center
\includegraphics[width = 0.329\textwidth]{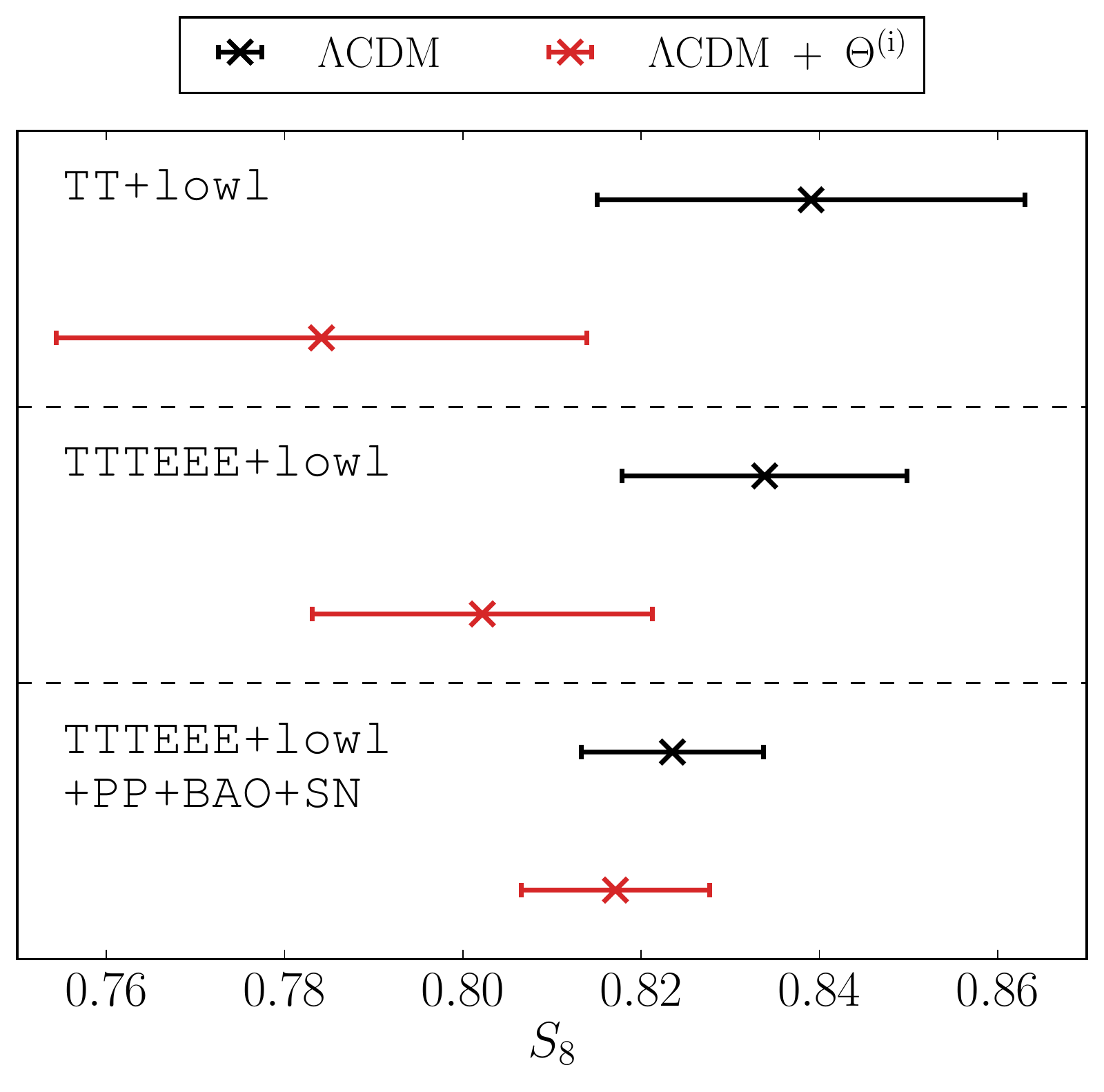}
\includegraphics[width = 0.329\textwidth]{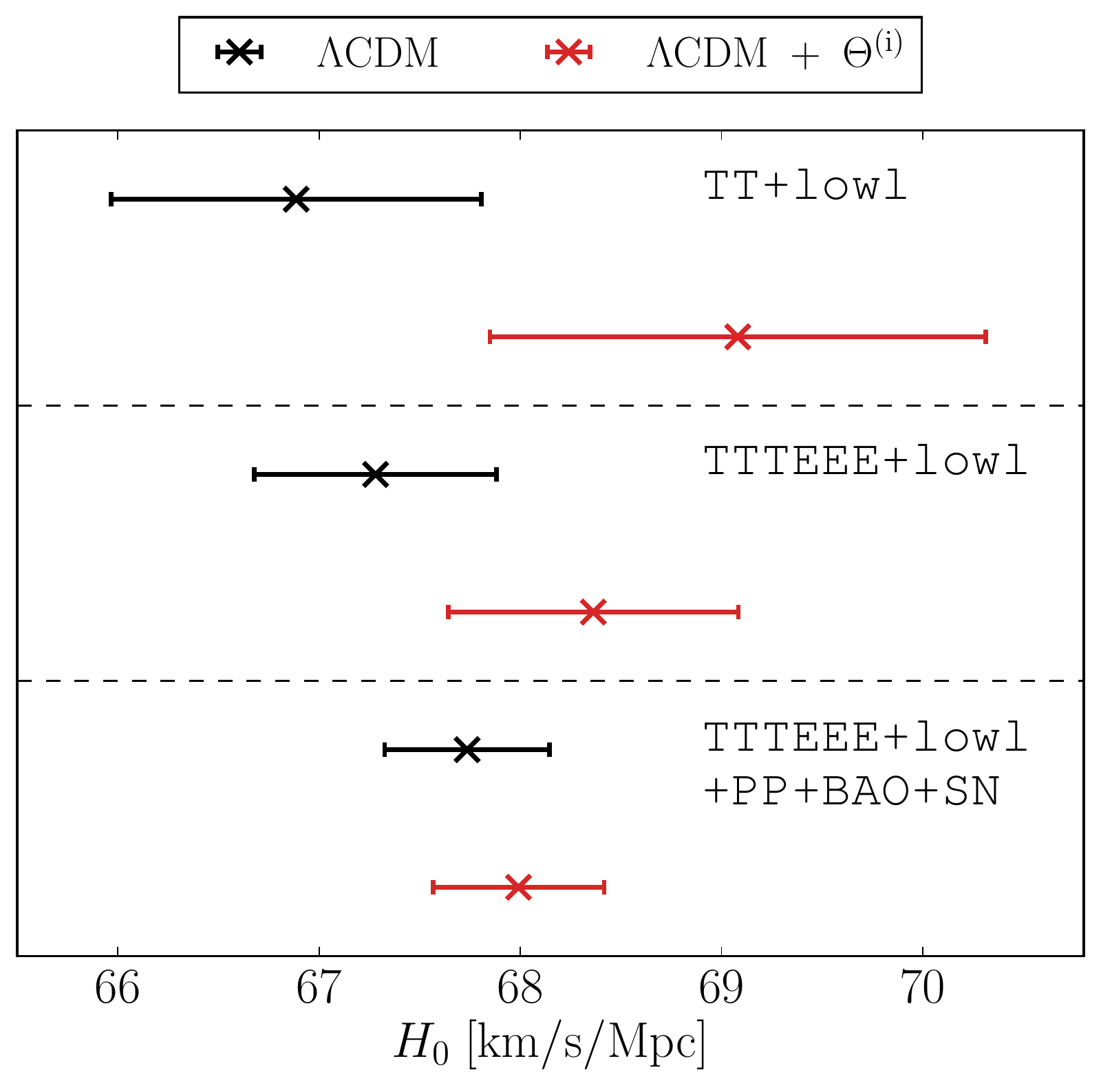}
\includegraphics[width = 0.329 \textwidth]{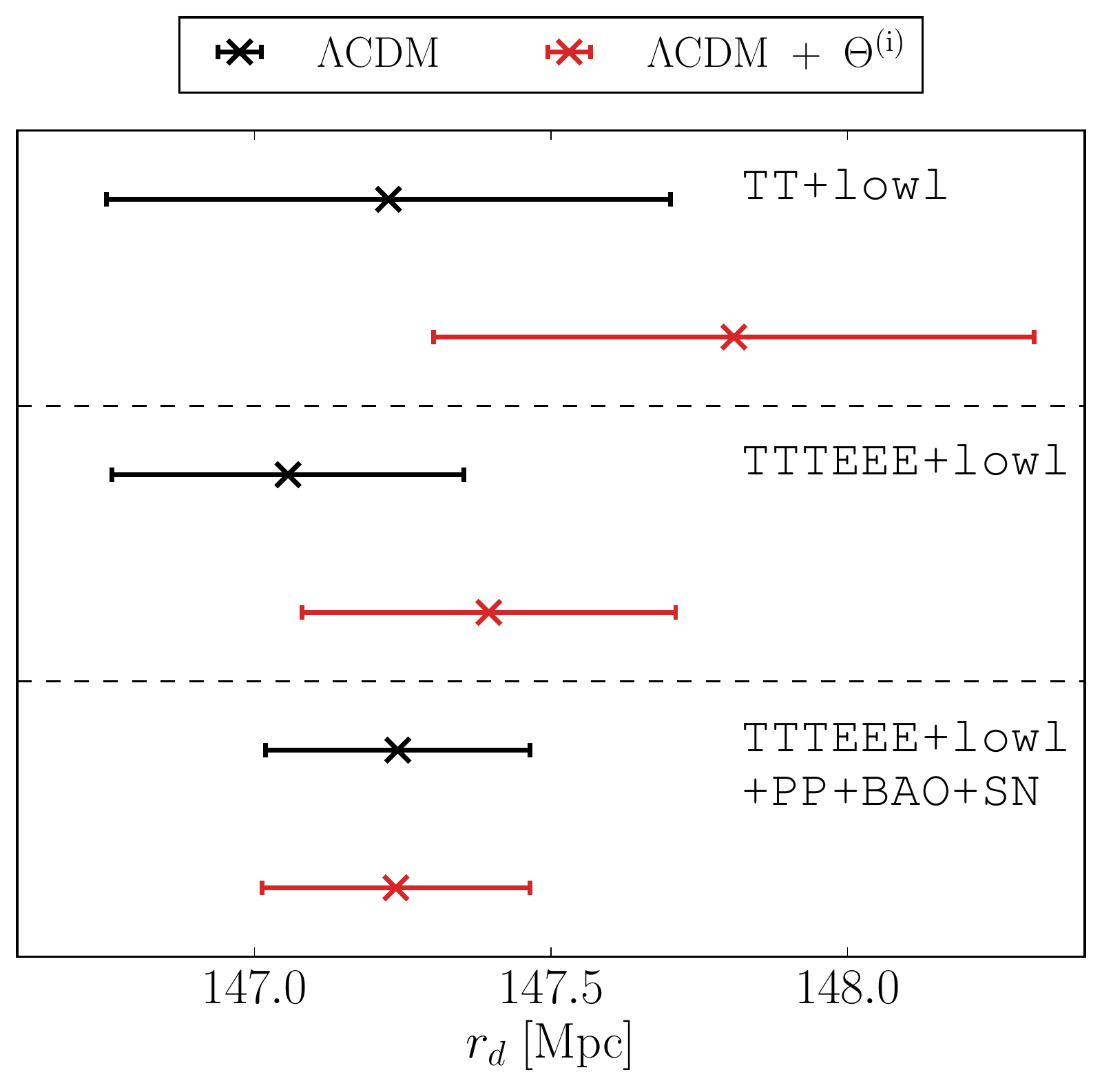}
\cprotect\caption{Impact of marginalizing over the lensing anomaly on $\Lambda$CDM tension parameters.
Constraints on $S_8$ {(left)}, $H_0$ {(center)} and $r_d$
{(right)} from \verb|TT+lowl| (top), \verb|TTTEEE+lowl| {(middle)} or
\verb|TTTEEE+lowl+PP+BAO+SN| (bottom) with (red) and without (black) marginalizing over
the lensing information contained in the smoothing of the peaks of
the temperature and polarization power spectra. 
}
\label{fig:lcdm}
\end{figure*}

In the top of Fig.~\ref{fig:lcdm}, we show changes to
constraints on the current value of the Hubble constant $H_0$ and the parameter $S_8 =
\sigma_8\sqrt{\Omega_m/0.3}$, which are the parameters involved in the most significant
tensions with other cosmological experiments. Here $\Omega_m = \Omega_c + \Omega_b$ and
$\sigma_8$ is the linear-theory root mean square fluctuations in spheres with a radius of
8 $h^{-1}$ Mpc at redshift $z = 0$. For reference, recent low-redshift measurements are
$H_0 = \(74.03 \pm 1.42\)\mathrm{km/s/Mpc}$ \cite{Riess:2019cxk} and $S_8 =
0.783^{+0.021}_{-0.025}$ \cite{Abbott:2017wau}.

For both parameters, the anomalously high lensing power in the Planck temperature and
polarization power spectra acts in the direction of increasing the
tension, although for $H_0$, this contribution is far from enough to explain majority of
the observed discrepancy. When using \verb|TTTEEE+lowl|, the parameters shift less, in
part due to preference of the polarization data for lower values of $H_0$ unrelated to
gravitational lensing \cite{Obied:2017tpd}.

As the $H_0$ tension can be restated in terms of disagreement in measurements of the comoving
size of the sound horizon to the drag epoch $r_d$ \cite{Bernal:2016gxb,Aylor:2018drw},
in Fig.~\ref{fig:lcdm} we  also show constraints on this parameter. Notice that
even though  $H_0$ increases when lensing is marginalized, $r_d$ and the sound
horizon get larger.  This is in the opposite direction than what is required for relieving
$H_0$ tension in the inverse distance ladder.  This is because the acoustic peak
positions still fix $\Omega_m h^3$ so that raising $H_0$ lowers $\Omega_m h^2$ (see
Fig.~\ref{fig:neff}).

Correspondingly, in the same Figure we see that  when we add  BAO,  supernova constraints,
and Planck lensing reconstruction,  which are the observables that are consistent with the
Planck best fit cosmology, the parameters shift back toward their $\Lambda$CDM values.
The  net shifts from marginalizing over the lensing anomaly end up being much smaller than the observed tensions. 

Overall, in the context of $\Lambda$CDM it is thus possible to conclude that better
understanding of the Planck lensing-like anomaly will not play an important role in
resolution of the observed tensions.

Assuming $\Lambda$CDM is the correct cosmological model, we can also revisit the
significance of the $\Theta^{(1)}$ tension discussed in the previous section. 
While previously we used only the ``unlensed'' part of \verb|TTTEEE+lowl| to
constrain the values of $\tilde \theta_A$, we can now fold in the \verb|PP| information,
to better constrain these $\tilde \theta_A$ parameters.
This allows us to make tighter predictions on $\Theta^{(i)}$ within the
context of $\Lambda$CDM. Compared with this improved $\Lambda$CDM prediction, the direct
measurement of $\Theta^{(1)}$ from the smoothing of the acoustic peaks is only 2.1$\sigma$
anomalous.  This is a fairer quantification of the Planck lensing-like tension {in $\Lambda$CDM} than just
looking at the internal tension in the temperature and polarization power spectra (or
$A_L$ in the usual approach), because now we are taking all of the Planck
data into consideration. Alleviation of the tension comes mostly from shifting
$\widetilde{\Omega_c h^2}$ back up to $0.1192 \pm 0.0012$, which in $\Lambda$CDM increases the
predicted lensing power.  Additionally, raising $\Omega_c h^2$ lowers radiation driving,
which smooths the acoustic peaks. This reduces the oscillatory residuals, lowering the
direct constraint on $\Theta^{(1)}$.

After adding \verb|BAO| and \verb|SN| data to improve the $\Lambda$CDM
prediction {further}, the tension goes mildly up, to 2.2$\sigma$. We do not use local
measurements of $H_0$ or weak lensing measurements in making the prediction because of the
aforementioned tensions.

\subsection{\LCDM + $N_\mathrm{eff}$}

Adding light relativistic degrees of freedom to the early Universe has been investigated
as a possible avenue to reduce the $H_0$ tension. Additionally,
constraints on $N_\mathrm{eff}$ are one of the main science drivers for the proposed CMB-S4
experiment \cite{Abazajian:2016yjj}. This motivates us to investigate how much 
does the Planck lensing-like anomaly affect parameter constraints in this model.

\begin{figure}
\center
\includegraphics[width = 0.49 \textwidth]{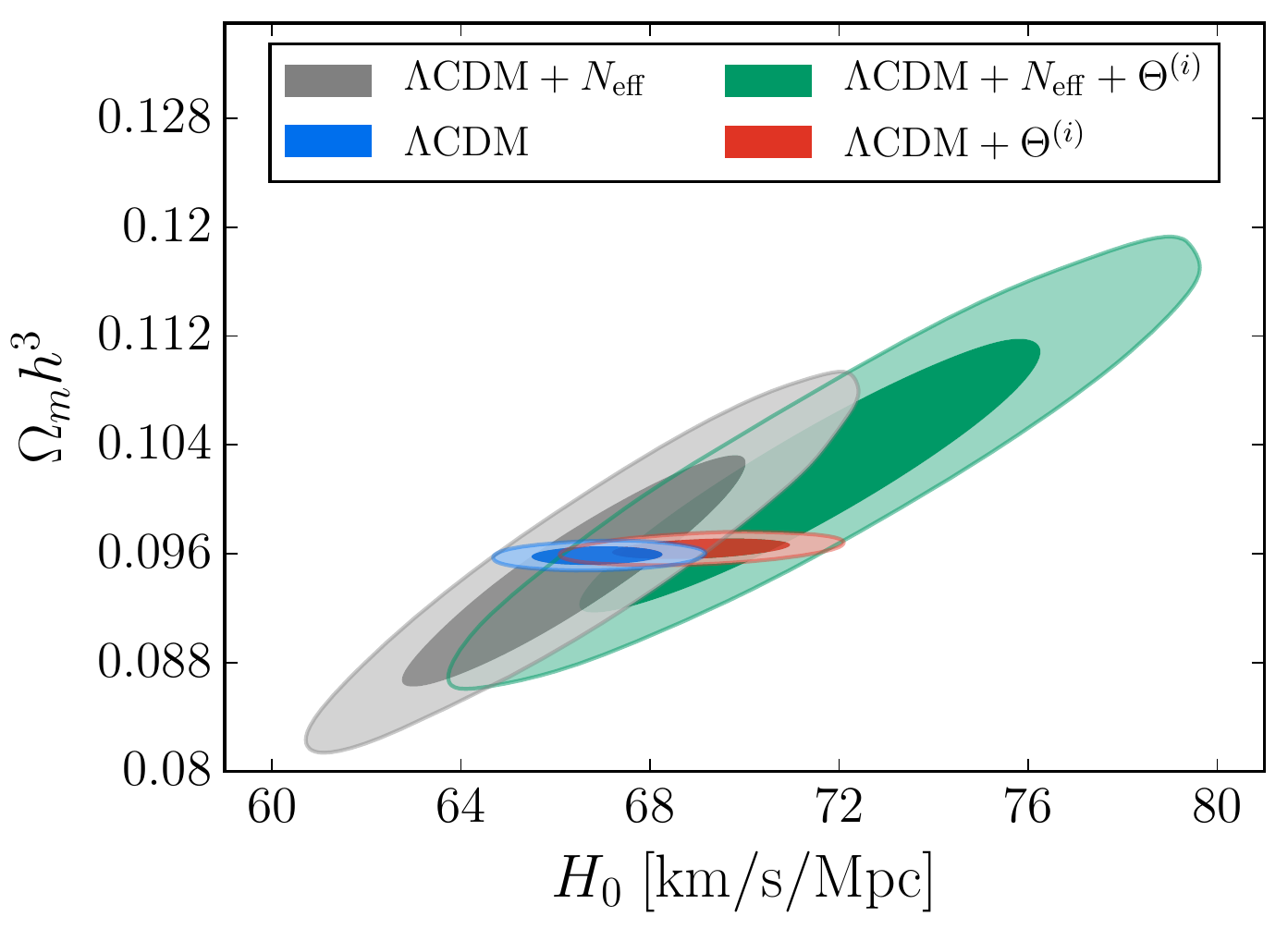}
\cprotect\caption{\verb|TT+lowl|  constraints on $\Omega_m h^3$ and $H_0$ in $\Lambda$CDM 
with (red) and without (blue) marginalizing over the lensing
anomaly. In green and gray respectively, we show the same constraints in
$\Lambda$CDM+$N_\mathrm{eff}$. 
}
\label{fig:neff}
\end{figure}

In Fig.~\ref{fig:neff}, we focus on \verb|TT+lowl| data and show constraints on $\Omega_m
h^3$ and $H_0$ in \LCDM and \LCDM+$N_\mathrm{eff}$, with and without marginalization over
the lensing PCs $\Theta^{(i)}$. It is visible that allowing arbitrary gravitational
lensing shifts the contour along the CMB $\Omega_m h^3$ degeneracy, while releasing $N_\mathrm{eff}$ from its fiducial value opens up
this degeneracy direction. When using only the \verb|TT+lowl| data, this combination of lensing marginalization and degeneracy
breaking allows an increase of $H_0$ to $\(71.5 \pm 3.2\)$ km/s/Mpc. However, as in $\Lambda$CDM, since lensing
mainly allows shifts at fixed $\Omega_m h^3$, 
adding \verb|BAO|, Pantheon supernovae and CMB lensing reconstruction
brings $H_0$ down to $\(68.3 \pm 1.4\)$ km/s/Mpc, because the corresponding constraints on $\Omega_m$ are
consistent with the Planck best fit values. For \verb|TTTEEE+lowl| the situation is
qualitatively similar, but already with \verb|TTTEEE+lowl| alone the
marginalization over $\Theta^{(i)}$ pushes $H_0$ to only $\(68.2 \pm 1.6\)$ km/s/Mpc.

\begin{figure}
\center
\includegraphics[width = 0.40 \textwidth]{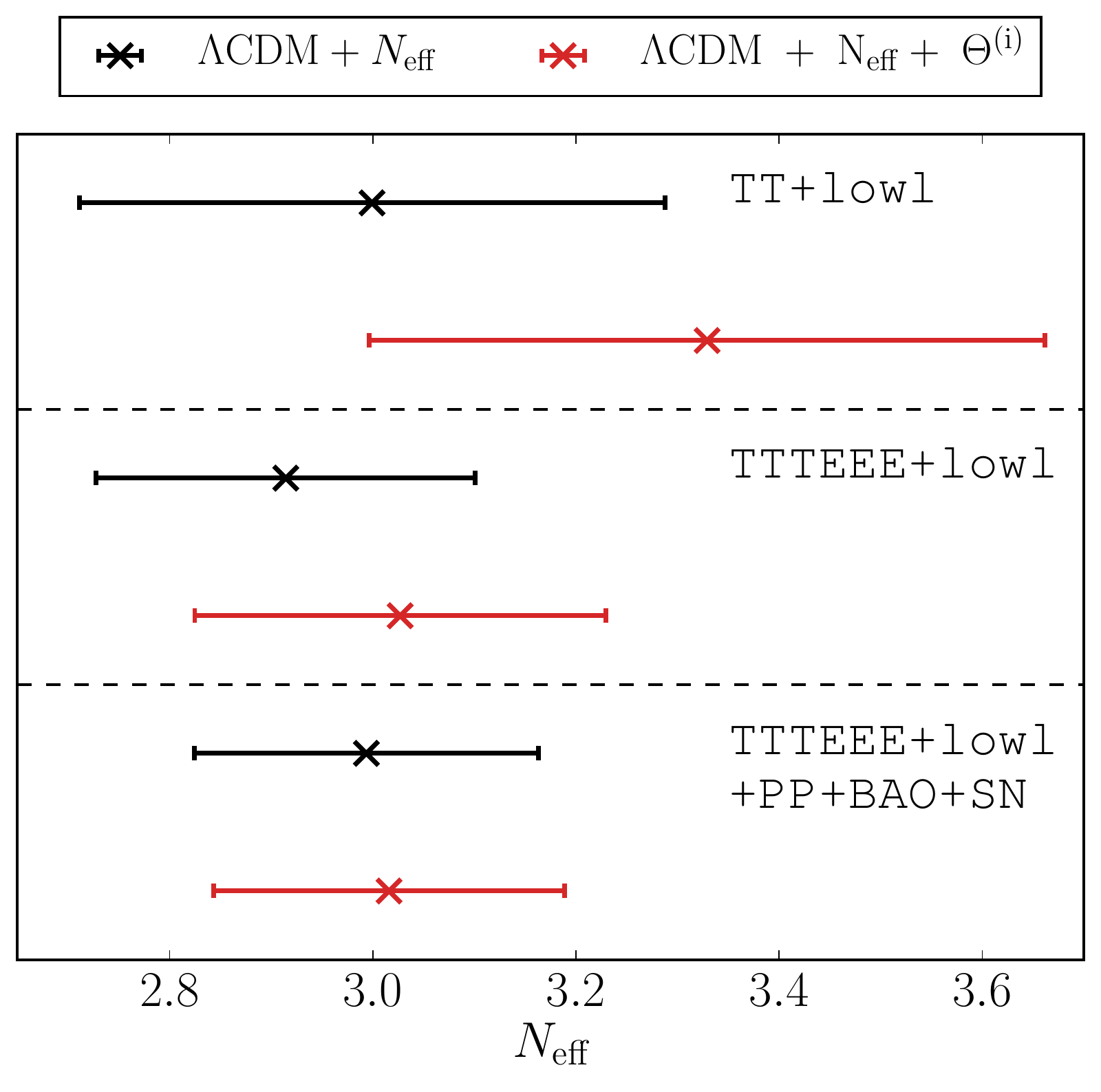}
\cprotect\caption{Impact of {marginalizing over} Planck lensing anomaly on $N_\mathrm{eff}$.  Constraints from \verb|TT+lowl| (top), \verb|TTTEEE+lowl| (middle) or
\verb|TTTEEE+lowl+PP+BAO+SN| (bottom) with (red) and without (black) marginalizing over
the lensing anomaly.
}
\label{fig:neff_bar}
\end{figure}

As for $N_\mathrm{eff}$ itself, we find its best fit value to be within one standard deviation
away from the standard model value 3.046 for all the cases we study both with and without
the lensing PCs $\Theta^{(i)}$ marginalized, see Fig.~\ref{fig:neff_bar}.

\subsection{\LCDM + $\sum m_\nu$}

\begin{figure}
\center
\includegraphics[width = 0.485 \textwidth]{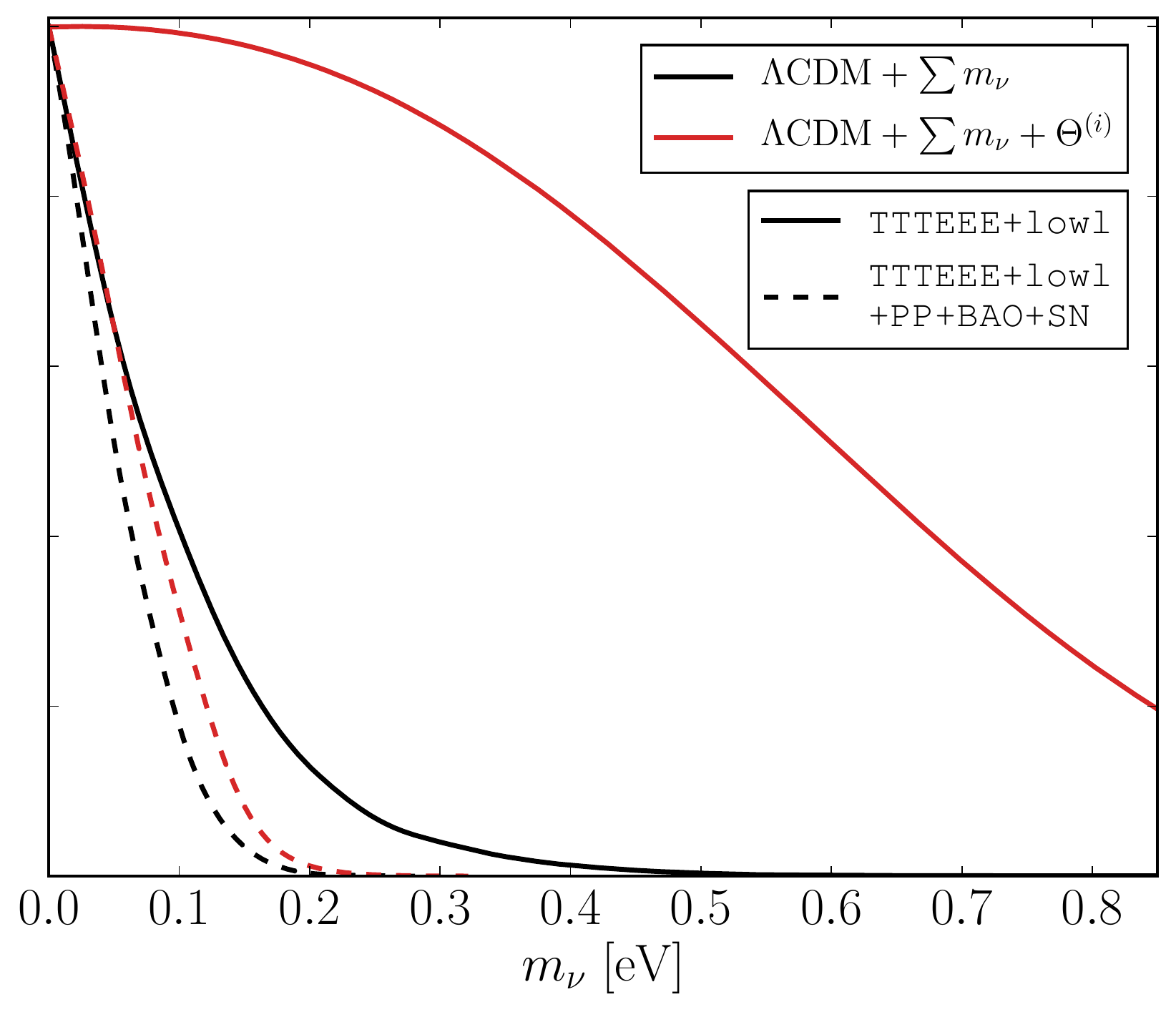}
\cprotect\caption{Constraints on $\sum m_\nu$ using
\verb|TTTEEE+lowl| (solid) and \verb|TTTEEE+lowl+PP+BAO+SN| (dashed) with (red) and
without (black)  marginalizing over the lensing anomaly.
}
\label{fig:mnu}
\end{figure}

When using only the CMB temperature and polarization power spectra, the most constraining
effect of neutrino masses is suppression of $C_L^\PP$ below the neutrino free-streaming
scale, which lowers the amount of smoothing of the acoustic peaks. With our methodology,
this is easy to see.  After marginalizing over $\Theta^{(i)}$, the constraints on $\sum
m_\nu$ degrade significantly, with the upper 95\% CL increasing from 0.57 eV to 1.1 eV
when using \verb|TT+lowl| and from 0.26~eV to 0.87~eV when using \verb|TTTEEE+lowl| (see
Fig.~\ref{fig:mnu} for the latter).

Conversely in $\Lambda$CDM {+$\sum m_\nu$},  the  lensing-like anomaly strengthens the constraints on neutrino
masses, as positive neutrino masses exacerbate the tension. To estimate importance of the
anomaly in this context, in Fig.~\ref{fig:mnu} we show constraints on $\sum m_\nu$ using
\verb|TTTEEE+lowl+PP+BAO+SN|, i.e. when additionally including Planck lensing
reconstruction, together with \verb|BAO| and \verb|SN| data.  We find that not
using the information contained in the amount of peak smoothing (by marginalizing over
$\Theta^{(i)}$) degrades the neutrino constraint by about 20\%, from the 95\% upper CL of
0.115~eV to 0.134~eV. 

The sum of the neutrino masses is well constrained even without the knowledge of the
amount of peak smoothing in the temperature and polarization power spectra. This is due to
different parameter degeneracy directions: the \verb|BAO+SN| data is sensitive to the sum
of CDM and neutrino energy densities, while to increase the lensing power in \verb|PP| it
is necessary to either increase $\Omega_c h^2$ or decrease the sum of the neutrino masses. 

\begin{figure}
\center
\includegraphics[width = 0.49 \textwidth]{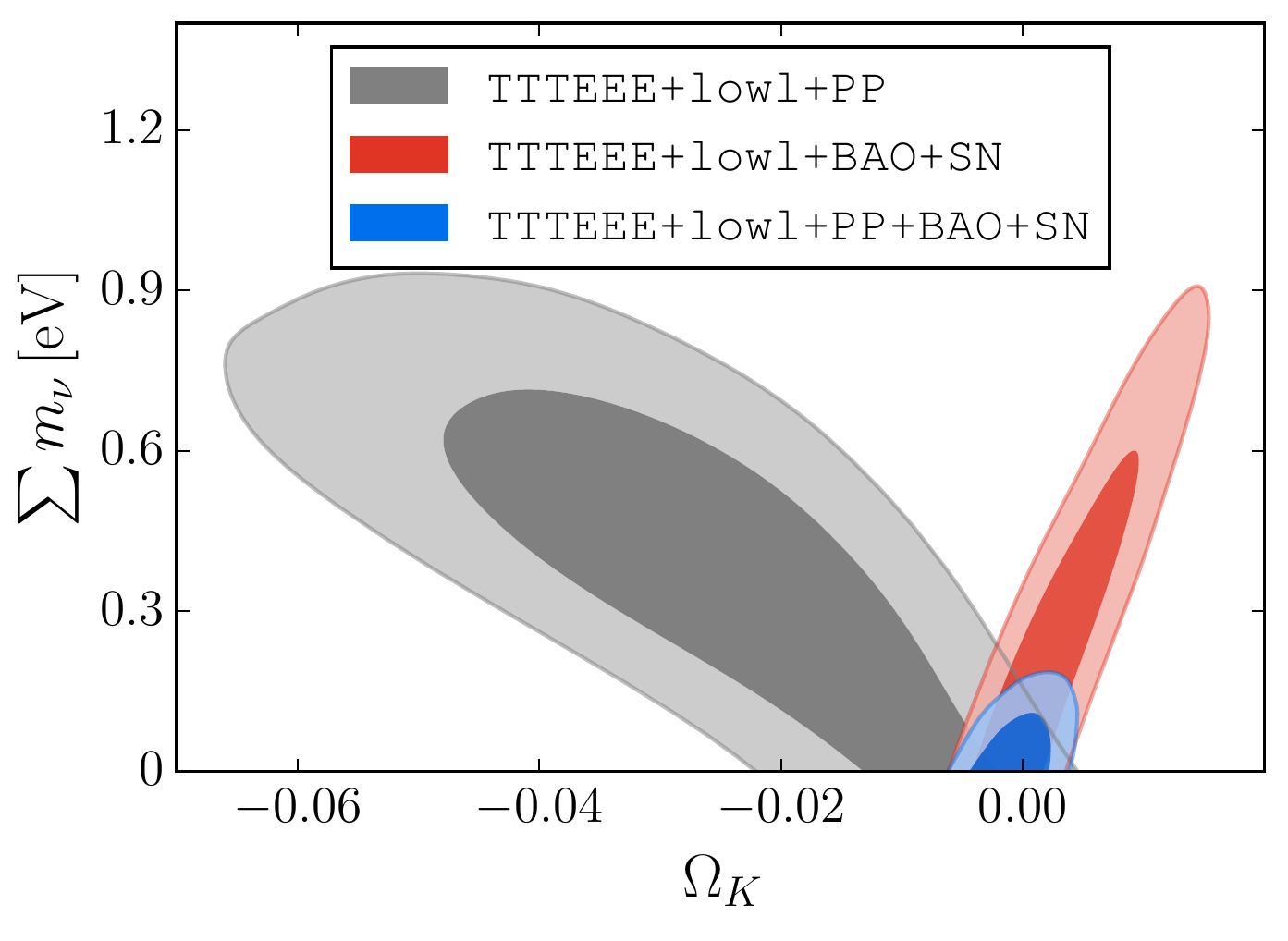}
\caption{
Constraints on the sum of the neutrino masses and the curvature density
parameter $\Omega_K$ after marginalizing the lensing anomaly in $\Lambda$CDM+$\Omega_K$+$\sum m_\nu$ + $\Theta^{(i)}$ for various
combinations of data sets.
}
\label{fig:omk}
\end{figure}

We also checked that in the full \verb|TTTEEE+lowl+PP+BAO+SN| data
set, considering non-flat cosmologies further degrades the constraint on $\sum m_\nu$ by
only about 10\%, again because of a degeneracy breaking (see Fig.~\ref{fig:omk}).
While the lensing anomaly can be translated into a preference for a closed Universe when
considering the temperature and power spectra alone, as recently emphasized by \cite{DiValentino:2019qzk}, we find
that adding either Planck lensing reconstruction or \verb|BAO+SN| leads to results
consistent with flat Universe. This preference is robust with respect to whether one
marginalizes over the lensing anomaly or
not.

\section{Discussion}
\label{sec:discuss}

In this work we studied constraints on CMB gravitational lensing potential from the final
release of the Planck satellite data. The results are consistent with our previous
analysis of an earlier release of the data. The biggest change in the final release are
improved constraints on the optical depth to recombination $\tau$, that {then improve} the
prediction of the lensing power expected within the $\Lambda$CDM model.

The overall picture remains the same: allowing for an arbitrary gravitational lensing
potential still leads to anomalously high lensing power determined from the smoothing of
the peaks in the temperature and polarization power spectra. This power is about 2$\sigma$
higher than the lensing power measured directly through the reconstruction of the gravitational
lensing potential and close to 3$\sigma$ higher when compared with the $\Lambda$CDM
prediction based on values of the standard cosmological parameters constrained only
through their effect on the unlensed power spectra (see Fig.~\ref{fig:1d}). The latter is
related to the usually quoted $A_L$ tension and is associated with the oscillatory residuals
in the temperature power spectrum.

Because these oscillatory residuals in conjunction with the low multipole 
anomalies also impact $\Lambda$CDM parameters in a manner that exacerbates tension \cite{Aghanim:2016sns,Obied:2017tpd}, we can make
a better $\Lambda$CDM prediction by combining parameter constraints
from Planck lensing reconstruction, \verb|BAO|, \verb|SN| and the ``unlensed'' portion of
\verb|TTTEEE+lowl| (marginalizing over the lensing information $\Theta^{(i)}$ in the CMB
power spectra). Compared with this joint prediction, the amount of the
gravitational lensing inferred from smoothing of the CMB acoustic peaks in Planck data is
only anomalous in $\Lambda$CDM at  2.2$\sigma$, comparable to the model independent value.

It should also be noted that when analyzing the Planck data using the
\verb|CamSpec| likelihood that is not currently publicly available, the significance of
the $A_L$ anomaly drops by approximately 0.5$\sigma$ \cite{Aghanim:2018eyx,
Efstathiou:2019mdh}. Repeating our analysis with the \verb|CamSpec| likelihood could thus
find a consistency at better than 2$\sigma$.

It is possible that such levels of discrepancy can be caused by a random
fluctuation. Even taken at face value, gravitational lensing is an unlikely cause of the
$A_L$ tension given that the two CMB sources of lensing information are discrepant for
any lens power spectrum and that 
 lensing reconstruction results are fully consistent with the
$\Lambda$CDM expectation. Furthermore, as both SPT temperature and $E$-polarization data show
preference for low amounts of lensing, it does not seem likely that physics beyond the
standard model that only affects the temperature and/or polarization power spectra is the
culprit, unless fine tuned to predominantly affect the range of scales where Planck is the
most sensitive to lensing effects.

Consequently, we also studied what impact marginalizing this lensing-like anomaly has on constraints of cosmological
parameters. When using only the temperature and large scale polarization Planck data,
 this reduces the tensions with the local
measurements of $H_0$ and $S_8$. However, additional data strongly restrict the
ability of this lensing-like anomaly to shift cosmological parameters within $\Lambda$CDM.
In this context, better understanding of Planck lensing-like anomaly is
thus not likely to shed more light on the $H_0$ and $S_8$ tensions. 

We also investigated $\Lambda$CDM+$N_\mathrm{eff}$ and $\Lambda$CDM+$\sum
m_\nu$, with similar conclusions. In $\Lambda$CDM+$N_\mathrm{eff}$, the extra freedom in
$N_\mathrm{eff}$ changes the relationship between $\Omega_m h^3$ and $H_0$ required
by  CMB
power spectra data, however marginalizing lensing mainly broadens the $H_0$ constraints at 
fixed $\Omega_m h^3$ and hence  constraints on $\Omega_m$ from data sets such as \verb|BAO|,
\verb|PP| and \verb|SN| drive the values of $H_0$ back to those found without marginalizing.

As is well known, the Planck lensing-like anomaly strengthens neutrino mass constraints.
When combining Planck data with current \verb|BAO| and \verb|SN| data, we find
that the lensing-like anomaly improves the neutrino mass constraints by less than 20\%.
Additionally allowing nonzero curvature further degrades this constraint
by only about 10\%. We find that when considering either \verb|PP| or \verb|BAO+SN| on top
of Planck temperature and polarization power spectra, the data are consistent with flat
Universe and this preference is not affected by the lensing anomaly.

In the future, additional data from experiments such as Simons Observatory and South Pole Telescope will improve our knowledge of the CMB, allowing us to better determine whether
this lensing-like anomaly is indeed just a statistical fluctuation or requires a more
fundamental explanation.

\acknowledgements{
We thank Tom Crawford, Cora Dvorkin, Silvia Galli, Marco Raveri, and Kimmy Wu for useful discussions.
WH was supported by U.S.~Dept.~of Energy contract DE-FG02-13ER41958 and the Simons Foundation.  
This work was completed in part with resources provided by the University of Chicago Research Computing Center.
}

\vfill
\appendix

\section{Tension in Multiple Dimensions}
\label{sec:appendix}

In the main paper, we quantify the tension 
between  the \verb|PP| and
\verb|TTTEEE+lowl| datasets
in a single parameter $\Theta^{(1)}$ with 3 other
$\Theta^{(i)}$ parameters marginalized.  In this Appendix we address the question of whether this
single parameter accurately captures the tension.

For the \verb|TTTEEE+lowl| dataset, the remaining parameters are partially or fully limited
by weak priors that allow a large volume of parameter space.  In this case their posterior probabilities
are highly non-Gaussian and can provide a misleading view of tension from difference in mean or related statistics.    Goodness of
fit statistics can also be misleading because the  \verb|TTTEEE+lowl|  best fit can lie in regions of parameter
space that are far from,  yet connected to,   models favored by $\verb|PP|$ and vice versa due to the weak priors.  

Since the  \verb|PP| 
dataset, and any analysis that includes it, is more constraining and its posteriors more Gaussian, we adopt the update difference
in mean technique introduced in~\cite{Raveri:2018wln}. 
Here we examine the difference in means between the  \verb|PP|  analysis and a
joint 
\verb|PP+TTTEEE+lowl|  analysis.

We isolate the linear combination of $\Theta^{(i)}$ parameters that can exhibit significant tension
using a Karhunen-Lo\`{e}ve decomposition of their covariance matrices
\cite{Raveri:2018wln}.  We find that only a single combination of $\Theta^{(i)}$ can exhibit such tension
 and it is dominated by $\Theta^{(1)}$, with a small $\Theta^{(2)}$ component. 
Moreover, the significance of the shift, 1.9$\sigma$, is in good agreement with the analysis 
in the main text.  This shows that marginalizing the additional $\Theta^{(i)}$ parameters
with weak priors that 
encompass large parameter volumes does not artificially lower the tension.

\vfill

\bibliography{plancklens_2019}

\end{document}